# Anisotropic Dzyaloshinskii-Moriya interaction protected by $D_{2d}$ crystal symmetry in two-dimensional ternary compounds

Yonglong Ga,[1,2] Qirui Cui,[1] Yingmei Zhu,[1] Dongxing Yu,[1] Liming Wang,[1] Jinghua Liang,[1] and Hongxin Yang[1,2,*]

[1]Ningbo Institute of Materials Technology and Engineering, Chinese Academy of Sciences, Ningbo 315201, China

[2]Center of Materials Science and Optoelectronics Engineering, University of Chinese Academy of Sciences, Beijing 100049, China

Correspondence: hongxin.yang@nimte.ac.cn

**ABSTRACT**

Magnetic skyrmions, topologically protected chiral spin swirling quasiparticles, have attracted great attention in fundamental physics and applications. Recently, the discovery of two-dimensional (2D) van der Waals (vdW) magnets have aroused great interests due to their appealing physical properties. Moreover, both experimental and theoretical works have revealed that isotropic Dzyaloshinskii-Moriya interaction (DMI) can be achieved in 2D magnets or ferromagnet-based heterostructures. However, 2D magnets with anisotropic DMI haven't been reported yet. Here, via using first-principles calculations, we unveil that anisotropic DMI protected by $D_{2d}$ crystal symmetry can exist in 2D ternary compounds $M$Cu$X_2$ ($M$: 3d transition metal (TM), $X$: group VIA). Interestingly, by using micromagnetic simulations, we demonstrate that ferromagnetic (FM) antiskyrmions, FM bimerons, antiferromagnetic (AFM) antiskyrmions and AFM bimerons can be realized in $M$Cu$X_2$ family. Our discovery opens up an avenue to creating antiskyrmions and bimerons with anisotropic DMI protected by $D_{2d}$ crystal symmetry in 2D magnets.

**INTRODUCTION**

Topological non-trivial magnetic structures such as chiral domain walls,[1] merons,[2] bimerons[3,4] and skyrmions[5,6] have attracted great research interests due to their rich physical properties and widespread application prospects in spintronic devices. Among these spin textures, magnetic skyrmions have been extensively studied due to their small size, low energy consumption and low driving current density.[7,8] Magnetic skyrmions with non-collinear spin configurations can be induced by antisymmetric exchange coupling-Dzyaloshinskii-Moriya interaction (DMI) in structures with inversion symmetry breaking and spin-orbit coupling (SOC).[9,10] The antisymmetric exchange interaction can be written as $E_{DMI} = \mathbf{D} \cdot (\mathbf{S}_i \times \mathbf{S}_j)$, where $\mathbf{D}$ is DMI vector and $\mathbf{S}_i$ and $\mathbf{S}_j$ represent spins of sites $i$ and $j$. Notably, magnetic skyrmions have been observed in non-centrosymmetric B20 bulk materials such as MnSi, FeGe, and FeCoSi[6,11-14] and interfacial systems of multilayers such as Ir(111)/Fe, Ta/CoFeB, and Pt/Co[15-21] with $C_{nv}$ crystal symmetry. On the other hand, anisotropic DMI are reported in ultrathin epitaxial Au/Co/W(110) with $C_{2v}$ crystal symmetry,[22] and antiskyrmions with anisotropic DMI are reported in acentric tetragonal Heusler compounds with $D_{2d}$ crystal symmetry[23,24] and non-centrosymmetric tetragonal structure with $S_4$ crystal symmetry.[25] In parallel with the development of hot study of skyrmions in these traditional bulk and multilayer thin films, 2D magnets, e.g., $Fe_3GeTe_2$,[26] $CrI_3$,[27] $CrGeTe_3$,[28] $MnSe_2$[29] and $VSe_2$[30] with long-range magnetic orderings have been extensively reported in the last few years, which have been providing an ideal platform to study fundamental properties of magnetism such as magneto-optical and magnetoelectric effect for ultracompact spintronics in reduced dimensions. Moreover, recent works have proposed that Néel-type skyrmions with isotropic DMI can be realized in 2D Janus magnets, e.g., Mn$XY$,[31] Cr$XY$[32] and multiferroics structures, e.g., CrN,[33] BaTiO$_3$/SrRuO$_3$,[34] In$_2$Se$_3$/MnBi$_2$Se$_2$Te$_2$.[35] However, it is worth noting that anisotropic DMI has not been reported yet in 2D magnets. Different from previous materials with isotropic DMI vector along with x and y directions, $M$Cu$X_2$ with special crystal symmetry has an anisotropic DMI vector. Skyrmion Hall effect will cause FM skyrmions with opposite topological charges to propagate in opposite direction, instead of moving parrel to injected current. Antiskyrmions Hall angle is strongly dependent on the direction of apply current related to the internal spin texture of antiskyrmion. When applying spin-polarized current to drive an antiskyrmion, the propagation direction of the antiskyrmion will follow the current direction without topological skyrmion Hall effect.[36-41] Therefore, it is possible to achieve the zero antiskyrmion Hall angle in a critical current direction.

In experiments, in order to discover 2D materials, a lot of efforts have devoted to finding materials with characteristics of weak interlayer bonds, which allow their exfoliation down to single layer by mechanical or liquid-phase approach[42,43] peeling off the three-dimensional layered van der Waals materials, such as CrGeTe$_3$,[28] CrI$_3$[27] and Fe$_3$GeTe$_2$[26] etc. Bulk FeCuTe$_2$ is a layered magnetic material stacked by weak interlayer van der Waals interaction, and it has been reported experimentally.[44-46] It is reported bulk FeCuTe$_2$ has a layered structure, and the unit cell parameters are as follows: $a$ = 3.93 Å, $c$ = 6.078 Å.[44] The layered compound FeCuTe$_2$ presents an antiferromagnetically ordered state below $T_N$ = 254 K.[47] Based on first-principles calculations, we further determine the magnetic properties of bulk FeCuTe$_2$. The calculated results show that the unit cell parameters $a$ and $c$ of bulk FeCuTe$_2$ are 3.964 Å and 6.176 Å, respectively. Moreover, we also compare energies with different magnetic orderings as shown in Supplementary Figure 10. Bulk FeCuTe$_2$ prefers AFM V in the ground state. From these results, one can see that theoretical results are consistent with experimental report. Before studying the properties of monolayer FeCuTe$_2$, we theoretically simulate the exfoliation energy when blocks were gradually peeled into a two-dimensional structure as shown in Supplementary Figure 1. The calculated cleavage energy is 0.46 J/m$^2$, which is comparable to that of graphene or phosphene,[48,49] indicating the high possibility that layered 2D FeCuTe$_2$ can be exfoliated from bulk. Interestingly, FeCuTe$_2$ with $D_{2d}$ crystal symmetry matches the Moriya symmetry rules[10] to achieve anisotropic DMI. Moreover, we predict a series of 2D $M$Cu$X_2$ structures with $D_{2d}$ crystal symmetry, where $M$ and $X$ represent the 3d transition metal (TM) and group VIA element, respectively. Finally, we realize a series of FM antiskyrmions, FM bimerons, AFM antiskyrmions and AFM bimerons structures.

**RESULTS AND DISCUSSION**

**Crystal structure of monolayer $M$Cu$X_2$**

Top and side views of the crystal structure of single-layer $M$Cu$X_2$ ($M$:3$d$ TM; $X$:group VIA ) are represented in Figs. 1(a)-(c). It exhibits a tetrahedral structure with $M$-$X_{top}$-$M$ and $M$-$X_{bot}$-$M$ configuration along x and y directions, and the face-centered coordinates are occupied by Cu atoms in the middle layer. The $M$Cu$X_2$ structure belonging to $D_{2d}$ (No.115) crystal symmetry has the symmetry generators: identity operation E: (x,y,z) → (x,y,z); mirror plane inversion M$_2$: (x,y,z) → (-x,-y,z); fourfold roto-inversion IC$_4$: (x,y,z) → (y,-x,-z); and twofold axis C$_2$: (x,y,z) → (x,-y,z), which leads to its feature with intrinsic inversion symmetry broken.

According to the Moriya rules[10], the induced DMI sign should be opposite along x and y directions.

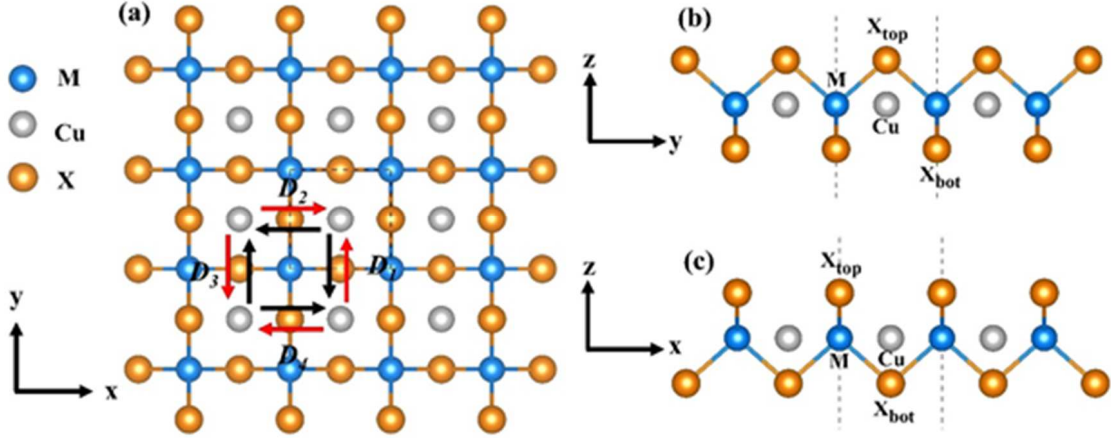

**Fig. 1 Crystal structure of *M*Cu*X*$_2$ monolayer, where *M* and *X* represent 3*d* TM and group VIA elements, respectively.** Top (a) and side (b-c) views of MCuX$_2$ monolayer. The dashed lines in (a-c) represent the unit cell. The red and black arrows in (a) indicate in-plane component of DMI with opposite chirality.

Via first-principles calculations, we obtain the basic magnetic parameters of *M*Cu*X*$_2$ ternary compounds as shown in Table 1 (Computation details are presented in experimental method). The $J_1$ and $J_2$ indicate the exchange coupling constants between NN and NNN atoms. In most systems, $J_1$ is orders larger than $J_2$. There are frustration coming from competing exchange interactions between $J_1$ and $J_2$ in classical Heisenberg model on the tetrahedra structure except for VCuS$_2$ and VCuSe$_2$. Interestingly, for *M*Cu*X*$_2$, when *M* changes from V to Mn and from Fe to Ni, the systems are tuned from FM to G-type AFM, respectively. In tetrahedra crystal, the five *d* orbitals of transition metal ions split into two groups lower *e* ($d_{x^2-y^2}$, $d_{z^2}$) and higher $t_2$ ($d_{xy}$, $d_{xz}$, $d_{yz}$) levels due to the influence of crystal field. The $t_2 \leftrightarrow p \leftrightarrow e$ super-exchange interaction favors to the appearance of FM coupling, while the $e \leftrightarrow e$ direct exchange and $t \leftrightarrow t$ direct exchange prefer to AFM coupling. When the $d$ orbit is more than half-filled with electrons, the AFM coupling mainly benefits from $t \leftrightarrow t$ direct exchange. Thus, in the $M$Cu$X_2$ family with $d$ orbit no less than half-filled, the competitive AFM coupling is stronger. Similar results are also depicted in the zinc-blende binary transition metal compounds.[50] Collecting all 3d TM atoms magnetic moments in $M$Cu$X_2$ monolayer [see Table 1]. We can find that Mn atoms have the highest magnetic moments and they monotonically decrease on both sides of Mn. It is obvious that the overall trend across 3d TM row obeys the Hund's rule.[51]

**DMI of monolayer $M$Cu$X_2$**

Figure 2 shows the calculated NN DMI of $M$Cu$X_2$ structures based on the chirality-dependent total energy difference approach.[52] It is found that all systems have anisotropic DMI and $d_x$ and $d_y$ have opposite signs along x and y directions, which is consistent with the DMI analysis at the beginning. Besides, the DMI strength varies from 0 to 15 meV/atom. These DMIs are very large compared to many state-of-the-art FM/HM heterostructures and 2D Janus structures, e.g., Co/Pt (~3.0 meV)[52] and Fe/Ir(111) (~1.7 meV)[53] thin films and 2D MnSTe (~2.63 meV).[31] In addition, it is worth noting that the DMI of VCuTe$_2$ reaches up to 15.2 meV/atom. In order to verify the correctness of the DMI, we calculated the variation of DMI when the U values of V atom were 2, 3, and 4 eV, which were 13.2, 15.2, and 11.6 meV/atom, respectively. Notably, CoCuTe$_2$ is different from other antiferromagnetic structures. We find that this structure tends to the Strip-AFM structure with different lattice constant in x and y direction. Meantime,

CoCuTe$_2$ also has the kind of anisotropic DMI in Fig. 2, which is -0.37 and 4.2 meV/atom along x and y directions, respectively.

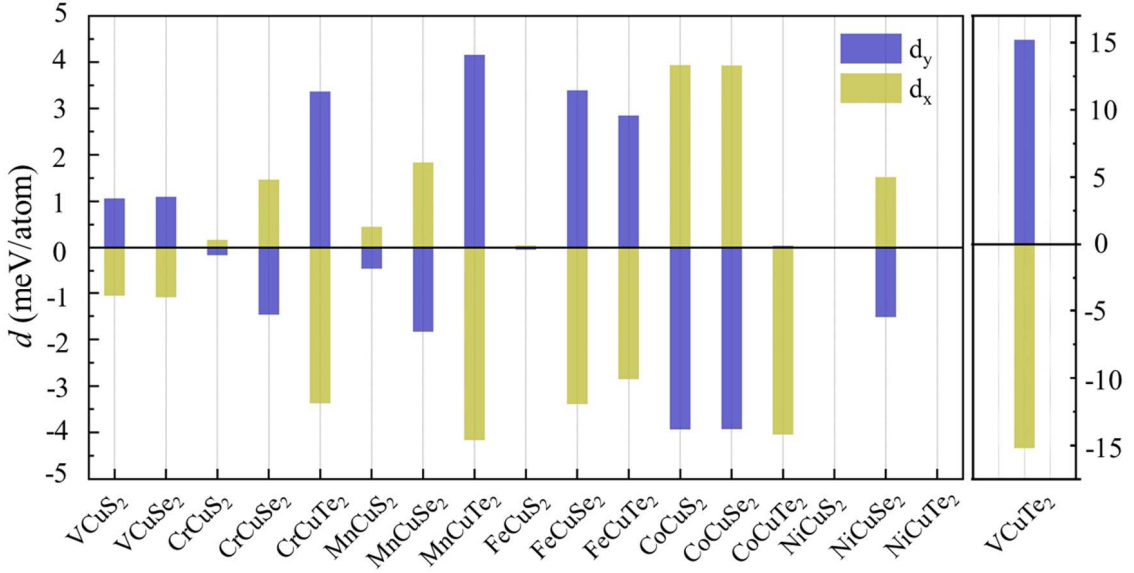

FIG. 2. **The calculated anisotropic DMI of monolayer MCuX$_2$.** Yellow and blue bars indicate the DMI components of NN magnetic atoms along x and y direction, respectively. Here, d > 0 (d < 0) represents the anclockwise (clockwise) chirality.

To elucidate microscopic energy source of DMI, we calculate the layer resolved SOC energy difference $\Delta E_{SOC}$ with opposite chirality associated with DMI. Only the $\Delta E_{SOC}$ are presented along the x direction in Figure 3. We can see that *M* atom and Cu atom contribute a relatively large DMI when the *X* is the light element S with weak SOC. As the *X* varies from S to Te, the dominant contribution of *X* element to DMI gradually improves in all *M*Cu*X*$_2$ monolayers due to the increase of SOC strength. Similar to the interface of HM/FM,[52,54] in which $\Delta E_{soc}$ is contributed mainly by heavy 5d metal elements of interfacial location. In our systems, Fert-Levy mechanism of DMI can be understood that heavy chalcogen element plays a significant role in inducing the spin-orbit scattering between two magnetic atoms. In addition, we noticed that the $\Delta E_{SOC}$ of NiCuTe$_2$ is close to zero. From optimization, we also identify the equilibrium lattice constant values for FM configuration of MnCuTe$_2$ and G-AFM configuration of NiCuTe$_2$, the calculated results are obtained about 4.100Å and 7.265Å, as shown in Supplementary Figure 3. The main reason is that the magnetic moment of Ni atoms in the system is very small, which leads to a small contribution to DMI. Of course, we also check the different U value from 2 to 4eV, a very small magnetic moments are obtained.

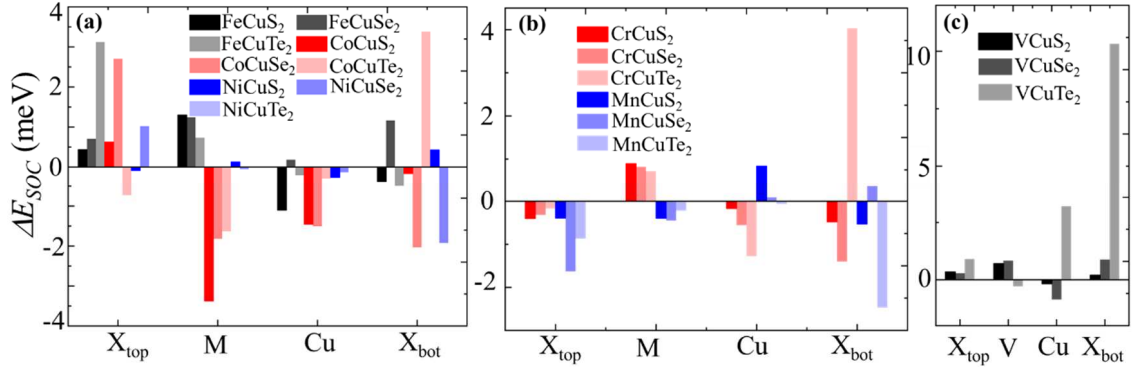

FIG. 3. **Layer-resolved DMI for MCuX$_2$ monolayers.** (a-c) Atom-resolved localization of SOC energy difference between clockwise and anticlockwise chirality for iron/cobalt/nickel (a), chromium/manganese (b) and vanadium ternary compounds.

Next, according to the Moriya's rules[10] and structural symmetry analysis above, the DMI vector for each pair of NNN $M$ atoms is parallel to their bonds because the 2-fold rotation axes are along the directions between two NNN magnetic atoms [see Supplementary Figure 2]. In our calculated structure with $D_{2d}$ crystal symmetry, the staggered spin vector will rotate in the plane perpendicular to the propagation direction <110>. However, we ignore the NNN DMI in our theoretical calculations, because we find that the DMI between NN and NNN atoms differs by about two orders of magnitude by using the four-state energy-mapping analysis,[55] e.g. VCuSe$_2$ and MnCuSe$_2$ (NNN DMI is -0.082meV and -0.075meV). Although the NNN DMI is neglected, we still observe the magnetic structure of Bloch-type helicoid from the results of the micromagnetic simulation.

**Chiral spin textures of monolayer *M*Cu*X*$_2$**

Furthermore, we perform the atomistic micromagnetic simulation based on first-principles calculated materials parameters are shown in Table 1 by using the VAMPIRE software package.[56] To get the dynamics of magnetization, the Landau-Lifshitz-Gilbert (LLG) equation was used with the Langevin dynamics as follows:

$$\frac{\partial \mathbf{S}_i}{\partial t} = -\frac{\gamma}{(1+\lambda^2)}\left[\mathbf{S}_i \times H_{eff}^i + \lambda \mathbf{S}_i \times \left(\mathbf{S}_i \times H_{eff}^i\right)\right], \qquad (1)$$

where $\mathbf{S}_i$ is the normal unit vector of *i*th magnetic atom, $\gamma$ is the gyromagnetic ratio and $\lambda$ is the gilbert damping constant. The magnitude of the effective field is obtained by the equation: $H_{eff}^i = -\frac{1}{\mu_i}\frac{\partial H}{\partial \mathbf{S}_i}$, in which $\mu_i$ represents magnetic moment of site *i* and $H$ is the Hamiltonian of the system. In all micromagnetic simulations, we relax a random state of 100-nm-wide with periodic boundary condition as the initial state to get the ground states without external magnetic field. For all FM systems, uniform FM and FM Néel chiral structures are observed from CrCuS$_2$ to MnCuTe$_2$ in Figure 4 due to the enhancement of ferromagnetic exchange coupling. Besides, in VCuSe$_2$ monolayer, antiskyrmions with diameter of 31 nm emerge on the large ferromagnetic domain. It is also important to know the topological charge Q. Here, we apply the lattice-based approach to calculate the topological charge. First, we perform the atomic spin model simulations to obtain the spin vector **S** on each lattice point. Furthermore, we use the formula: $\mathbf{S} \cdot (\partial_x \mathbf{S} \times \partial_y \mathbf{S})$ to calculate the topological charge density of each lattice point. Finally, we integrate the charge density of zone holding the topological magnetic quasiparticle to obtain the final topological charge. The topological charge is +1 in the zoomed antiskyrmion in Figure 4(d). More interestingly, we also realize the FM anti-bimerons arising from strong in-plane magnetic anisotropy in MnCuTe$_2$. In all AFM systems, the topological charge of each AFM antiskyrmion with antiparallel NN spin alignment is 0. It can be decomposed into two identical FM antiskyrmion sublattices, where the FM sublattice pairs have opposite topological charges +1 and -1. In FeCuSe$_2$, AFM antiskyrmion and AFM antiskyrmionium are presented in Figures 4(i)-(j). It is well known that lots of topologically magnetic textures can be induced by exchange frustration.[57,58] Notably, previous work have demonstrated that DMI and exchange frustration can stabilize skyrmionium in CrGe(Se,Te)$_3$ Janus monolayer when $J_1/J_2$ is within a certain range.[59] Similar to AFM antiskyrmion, AFM antiskyrmionioum with zero topological charge refers to a magnetic texture that can be view as two nested AFM antiskyrmions due to the frustration caused by opposite sign exchange coupling $J_1$ and $J_2$. In addition, we note that isolated bimerons and multiple bimerons are observed in CoCuS$_2$ and MnCuTe$_2$, respectively. The main reason is that FM exchange coupling can be consistent with an effective FM external field. When we reduce the $J_1$ of CoCuS$_2$, multiple bimerons appear, as is shown in Supplementary Figure 9. Similar to that CoCuS$_2$, when we increase $J_1$ or external magnetic field in the VCuS2 system, respectively, one can observe that isolated antiskyrmions emerge, as is shown in Supplementary Figure 11. Meantime, we also adopt open boundary condition to simulate the spin textures of FM MnCuSe$_2$ and AFM CoCuSe$_2$, as shown in Supplementary Figure 6. One can see that the simulated FM antiskyrmions in MnCuSe$_2$ and AFM antiskyrmions in CoCuSe$_2$ under open boundary condition have almost same size and topological property with periodic boundary condition. Furthermore, we also take into account the effect of dipolar interaction on chiral magnetic textures. For antiferromagnetic systems, magnetic dipolar interaction diminishes due to the cancellation of magnetic moment of coupled sublattices.[60] Thus, we simulated $VCuS_2$ and $MnCuSe_2$ monolayers based on first-principles calculated magnetic parameters, one can see that the calculated spin configurations of $VCuS_2$ and $MnCuTe_2$ are consistent with previous results as shown in the Supplementary Figure 7.

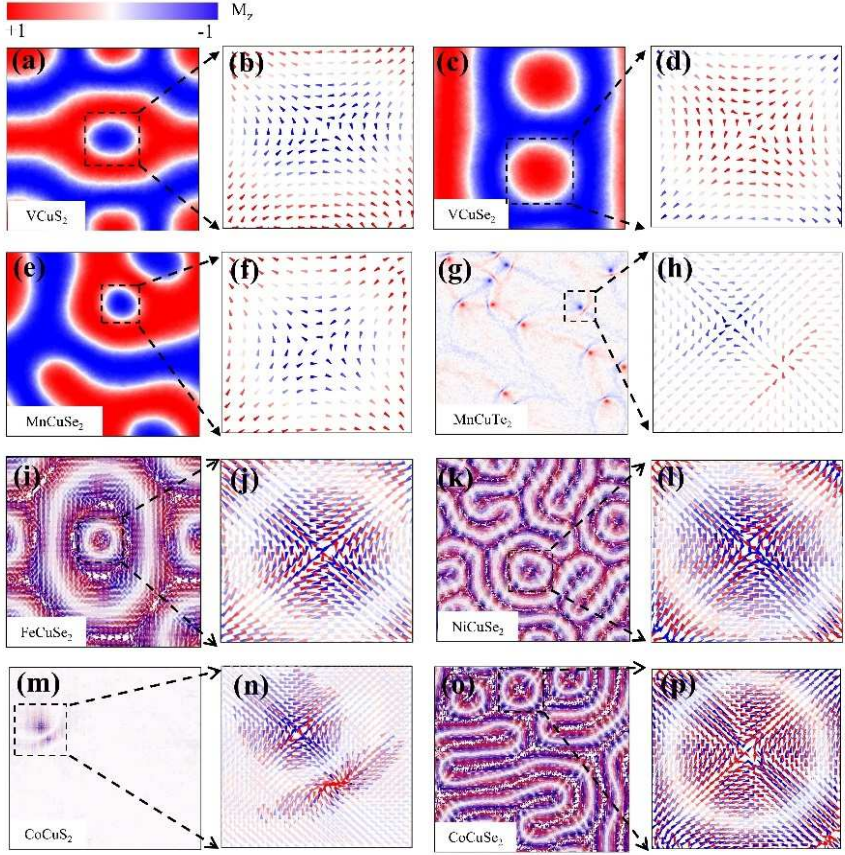

FIG. 4. **Spin configurations of ground states of monolayers MCuX$_2$ in real-space**. Spin configuration of a 100×100 nm square and zoom of isolated chiral magnetic profile of VCuS$_2$ (a-b) ; VCuSe$_2$ (c-d) ; MnCuS$_2$ (e-f) ; MnCuTe$_2$ (g-h) ; FeCuSe$_2$ (i-j); NiCuSe$_2$ (k-l); CoCuS$_2$ (m-n) and CoCuSe$_2$ (o-p). The color map indicates the out-of-plane spin component.

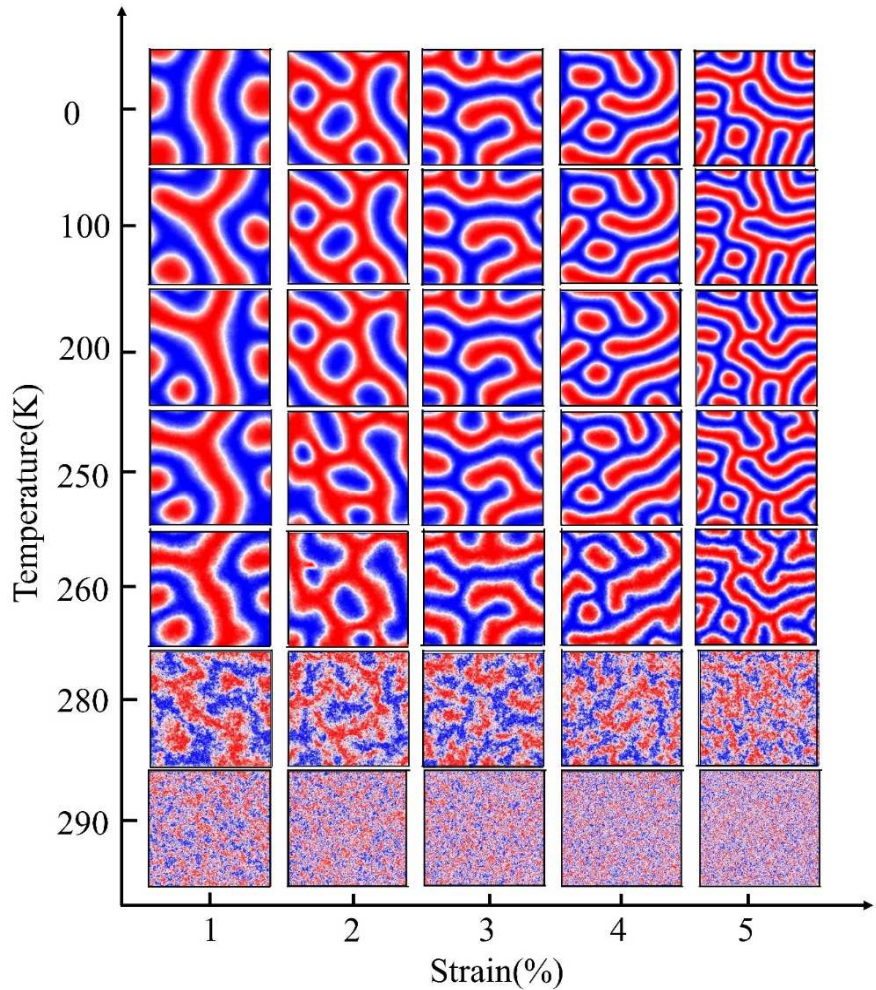

FIG. 5. **Phase of spin configurations under increased tensile strain and decreased temperature for $VCuSe_2$ monolayer.**

In addition, we also calculate magnetic parameters $J_1$, $J_2$, $K$ and $D$ of monolayer $VCuSe_2$ when the tensile strain increases from 1% to 5%, as shown in Table 2. We find that NN and NNN FM exchange coupling strengths decrease a lot while DMI changes slightly, resulting in the large ratio of $D/J$. Therefore, antiskyrmion with smaller diameter is achieved under tensile strain.[7] The phase diagram of VCuSe$_2$ monolayer under different stress and temperature is shown in Figure 5. For pristine VCuSe$_2$, the FM antiskyrmions embedded in the background of large-size domain are observed [Figure 4(c)]. It is found that stable chiral domain and magnetic antiskyrmions appear slowly when the temperature varies from 300 K to 250 K. Furthermore, if we keep the temperature decreasing, one can see that the chiral magnetic structures can be basically stabilized at 250K, and these chiral structures also undergo some small changes as the temperature continues to decrease to the finite temperature. Moreover, as the strain increases, the density of antiskyrmions increases and the size of domain becomes much smaller than that of pristine state. Supplementary Figure 8 presents the results of the micromagnetic simulations for spin spiral length as a function of strain. We can clearly observe the strain dependence of spin spiral length with the increasing $D$ and decreasing exchange $J$, which is consistent with $|D/J|$.

In summary, we discover a group of 2D layered ternary compounds $M$Cu$X_2$ ($M$:3d TM; $X$:group VIA) with anisotropic DMI protected by $D_{2d}$ crystal symmetry from first-principles calculations. We show that the anisotropic DMI can vary from 0 to 15 meV/atom in 2D layered ternary compounds $M$Cu$X_2$, where $M$ represents 3d transition metal, and $X$ represents the VIA group element. Thanks to the large enough anisotropic DMI, we demonstrate that a series of FM (AFM) antiskyrmions, bimerons and antiskyrmionioum can be realized without external field in the $M$Cu$X_2$ family. The discovery will provide a platform to find various FM/AFM anti-topological spin textures with crystal symmetry protected anisotropic DMI. In addition, our calculations show that magnetic parameters of VCuSe$_2$ are sensible to strain, and the possibility of antiskyrmions formation up to hundred Kelvin is demonstrated in 2D VCuSe$_2$. Our work will benefit both fundamental research and applications in the fields of 2D van der Waals materials and spintronics.

## METHODS

**DFT calculations**

First-principles calculations are carried out based on density functional theory (DFT) implemented in Vienna ab-initio Simulation Package (VASP).[61] We adopt Perdew-Burke-Ernzerhof (PBE) functionals of the generalized gradient approximation (GGA).[62] as the exchange correlation potential, and use projector augmented plane wave (PAW) method[63,64] to deal with the interaction between nuclear electrons and valence electrons. We set the cutoff energy of 520 eV for the plane wave basis set, and 24×24×1 k-point with Γ-centered meshes for the Brillouin zone integration. Partially occupied d orbitals of transition metal atoms are treated by GGA+U[65] with U = 3 eV for the 3d orbitals of *M* and Cu elements. We set a vacuum layer with thickness of 25Å in the z direction to ensure that there is no interaction between the periodic images. The convergence criteria of the total energy in the ion relaxation process and the Hellmann-Feynman force between atoms were set to be 1×10$^{-7}$eV and 0.001eV/Å, respectively. To describe our magnetic system, we adopt the following Hamiltonian model:

$$H = -J_1 \sum_{\langle i,j \rangle} \mathbf{S}_i \cdot \mathbf{S}_j - J_2 \sum_{\langle i,j \rangle} \mathbf{S}_i \cdot \mathbf{S}_j - \sum_{\langle i,j \rangle} \mathbf{D}_{ij} \cdot (\mathbf{S}_i \cdot \mathbf{S}_j) - K \sum_i (\mathbf{S}_i^z)^2, \quad (2)$$

where **S**$_i$ (**S**$_j$) is the normal spin vector of *i*th (*j*th) magnetic atom, the $J_1$ and $J_2$ represent exchange coupling constants between Nearest-Neighbor (NN) and Next-Nearest-Neighbor (NNN) atoms, respectively. *K* is magnetic anisotropy constant and $\mathbf{D}_{ij}$ is the DMI vectors. The methods to calculate *J*, *K*, *D* is described in the experimental section.

**Magnetic parameters**

*Exchange coupling constant*: we construct a 2×2×1 supercell to study three different magnetic configurations, which are FM, G-type AFM where nearest-neighbor spins are aligned antiparallel and Stripe-type AFM where spins are ordered antiferromagnetically (ferromagnetically) along with x (y) axis (see Supplementary Figure 4). Exchange coupling constant of nearest neighbor (NN) and next nearest neighbor (NNN) magnetic atoms are obtained based on the following formula:

$$E_{FM} = -\frac{1}{2} * 4(4J_1 + 4J_2) + E_0 \quad (3)$$

$$E_{AFM\,I} = -\frac{1}{2} * 4(-4J_2) + E_0 \quad (4)$$

$$E_{AFM\,II} = -\frac{1}{2} * 4(-4J_1 + 4J_2) + E_0 \quad (5)$$

$$J_1 = \frac{E_{AFM\,II} - E_{FM}}{16} \quad (6)$$

$$J_2 = \frac{2 * E_{AFM\,I} - E_{AFM\,II} - E_{FM}}{32} \quad (7)$$

where the positive/negative value corresponds to FM/AFM coupling.

*Magnetic anisotropy energy K*: magnetic anisotropy energy is defined as the energy difference between in-plane magnetized [100] axis and out-of-plane magnetized [001] axis:

$$K = E_{100} - E_{001} \quad (8)$$

*NN Dzyaloshinskii-Moriya interaction* (NN-DMI) *D*: we performed DMI calculations using the chirality-dependent total energy difference method.[52] First of all, a 4×1×1 supercell is constructed to obtain the charge distribution of system's ground state by solving the Kohn - Sham equations in the absence spin orbit coupling (SOC). Then, SOC is included and we set spin spirals to determine the self-consistent total energies in the clockwise and anticlockwise rotation. Finally, the energy difference between clockwise and anticlockwise rotation is calculated to obtain the anisotropic *D*. The DMI can be obtain by following formula:

$$D = (E_{cw} - E_{acw})/8 \qquad (9)$$

*NNN Dzyaloshinskii-Moriya interaction* (NNN-DMI): we performed the four-state energy mapping analysis.[55] First of all, a $4 \times 4 \times 1$ supercell is constructed to set all the spin configuration in the y direction, then change the spins between the two NN atoms. The DMI between NN *M* spins were calculated based on four spin configurations: (i) $\mathbf{S}_1 = (S,0,0)$, $\mathbf{S}_2 = (0,0,S)$; (ii) $\mathbf{S}_1 = (S,0,0)$, $\mathbf{S}_2 = (0,0,-S)$; (iii) $\mathbf{S}_1 = (-S,0,0)$, $\mathbf{S}_2 = (0,0,S)$; (iv) $\mathbf{S}_1 = (-S,0,0)$, $\mathbf{S}_2 = (0,0,-S)$. Next, according to spin interaction energy based different spin configurations, we can solve the in-plane component $D_y$: $D_y = (E_1 + E_4 - E_2 - E_3)/(4\mathbf{S}^2)$.

**Phonon spectrum**

In calculations, based on the PHONOPY code,[66,67] we use $3 \times 3 \times 1$, $3\sqrt{2} \times 3\sqrt{2} \times 1$ and $4 \times 4 \times 1$ supercells with the frozen phonon approximation to calculate the phonon dispersions of single layer *M*Cu*X*$_2$. Supplementary Figure 5 shows that $\Gamma$ point of some of *M*Cu*X*$_2$ monolayers have very small imaginary frequencies within the entire wave-vector space, which can be attributed to wavelength of particular mode.[68]

**DATA AVAILABILITY**

The datasets used in this article are available from the corresponding author upon request.

**CODE AVAILABILITY**

Code that supports the findings of this study are available from the corresponding authors on reasonable request.

**ACKNOWLEDGMENTS**

This work was supported by National Natural Science Foundation of China (Grant Nos. 11874059 and 12174405); Key Research Program of Frontier Sciences, CAS (Grant NO. ZDBS-LY-7021); Pioneer and Leading Goose R&D Program of Zhejiang Province (Grant No. 2022C01053); Ningbo Key Scientific and Technological Project (Grant No. 2021000215); Zhejiang Provincial Natural Science Foundation (Grant No. LR19A040002); Beijing National Laboratory for Condensed Matter Physics (Grant No. 2021000123); and Ningbo 3315 project

**AUTHOR CONTRIBUTIONS**

H.Y. conceived the project. Y.G. performed the calculations. Y.G., Q.C. and H.Y. analysis the results and wrote the manuscript. All the authors contributed to the discussion and the writing.

**COMPETING INTERESTS**

The authors declare no competing interests.

# Supplementary information

# Anisotropic Dzyaloshinskii-Moriya interaction protected by $D_{2d}$ crystal symmetry in two-dimensional ternary compounds

Yonglong Ga,[1,2] Qirui Cui,[1] Yingmei Zhu,[1] Dongxing Yu,[1] Liming Wang,[1] Jinghua Liang,[1] and Hongxin Yang[1,2,*]

[1]Ningbo Institute of Materials Technology and Engineering, Chinese Academy of Sciences, Ningbo 315201, China

[2]Center of Materials Science and Optoelectronics Engineering, University of Chinese Academy of Sciences, Beijing 100049, China

Correspondence: hongxin.yang@nimte.ac.cn

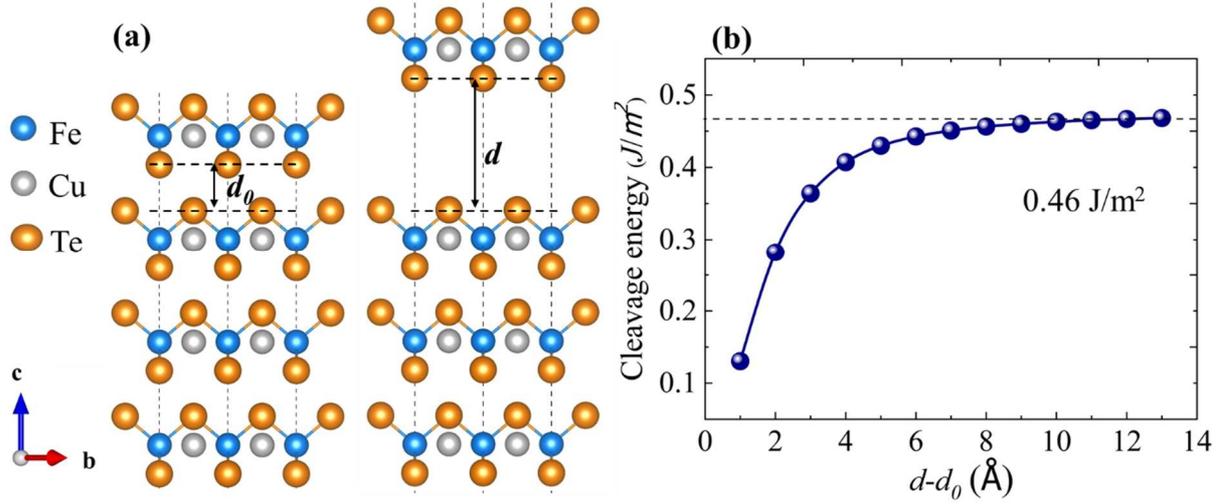

Supplementary Fig. 1. Cleavage energy estimation for the formation of monolayer FeCuTe$_2$. (a) Schematic illustration of the exfoliation procedure. (b) Cleavage energy as a function of $d-d_0$. Here, $d_0$ represents the equilibrium separation and $d$ represents the distance of the cleaved layer from the surface.

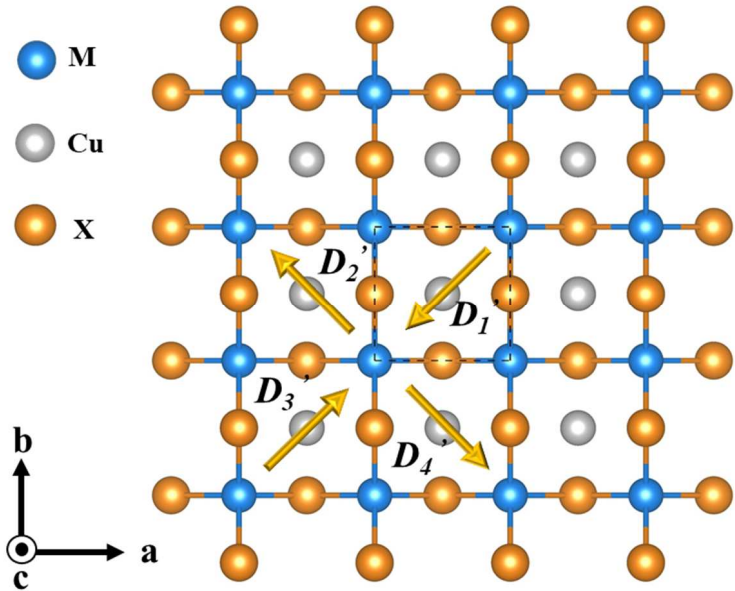

Supplementary Fig. 2. Top view of $M$Cu$X_2$ monolayers, where $M$ and $X$ represent the TM and group VIA elements, respectively. The orange arrows represent in-plane component of NNN DMI.

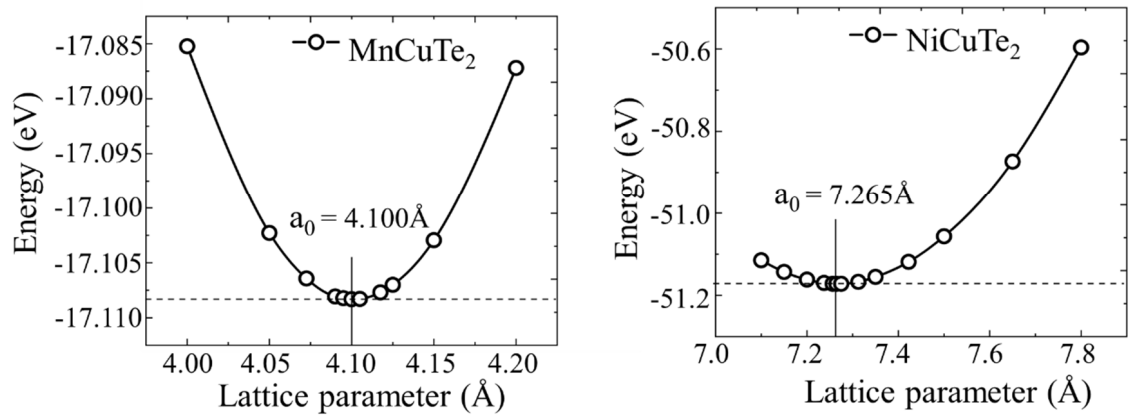

Supplementary Fig. 3. Energy convergence test to obtain the lattice parameters of MnCuTe$_2$ and NiCuTe$_2$.

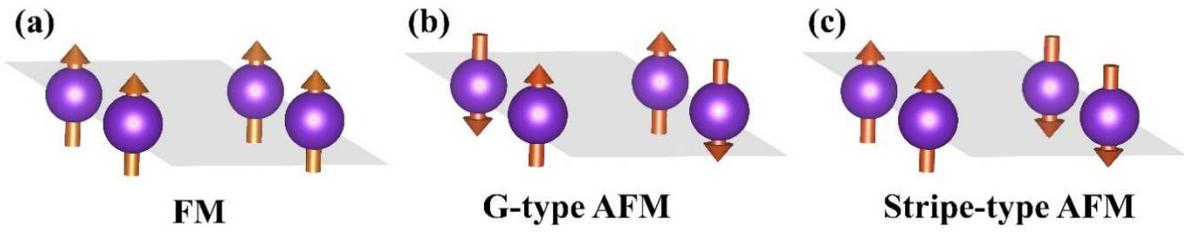

Supplementary Fig. 4. Different magnetic configurations for monolayer $M$Cu$X_2$. Including (a) ferromagnetic (FM), (b) G-type antiferromagnetic (G-type AFM) and (c) Stripe-type antiferromagnetic (Stripe-type AFM) configurations.

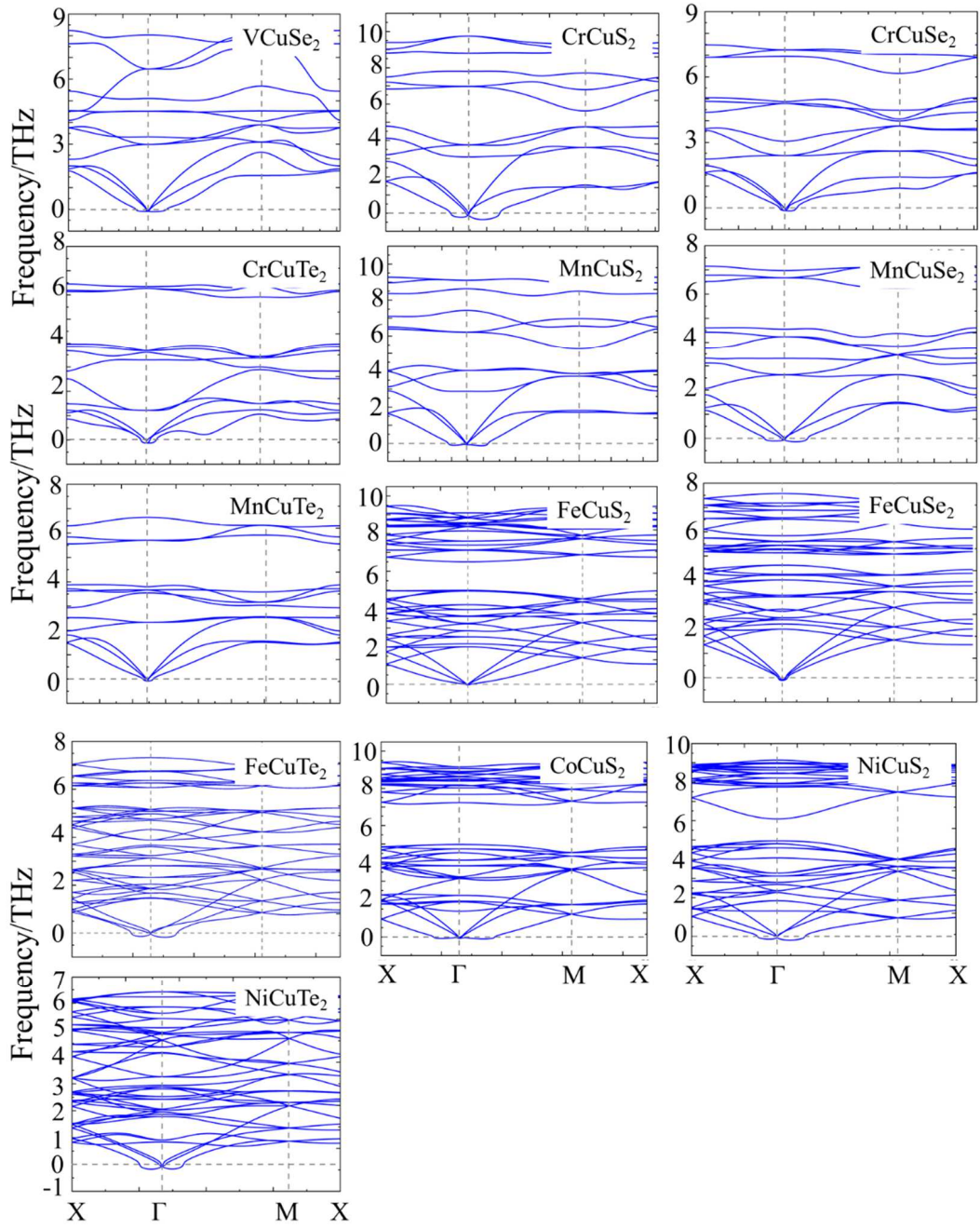

Supplementary Fig. 5. The phonon spectra for *M*Cu*X*$_2$ monolayers.

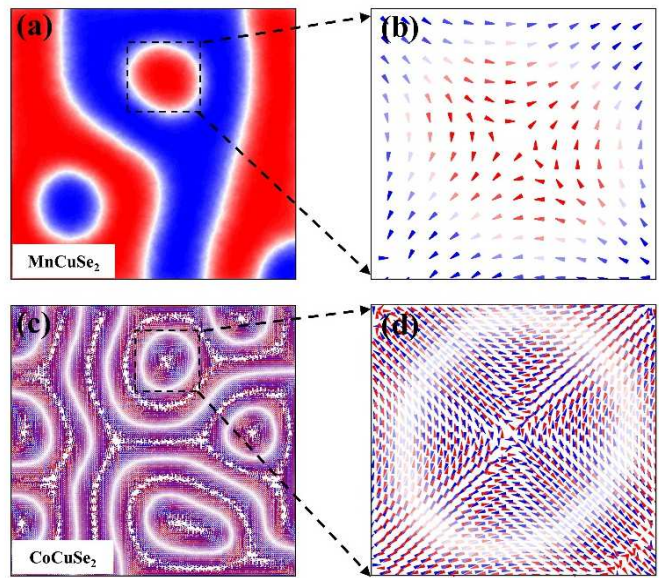

Supplementary Fig. 6. The spin configurations of 100 nm square and zoom of isolated magnetic profile of MnCuSe$_2$ (a), CoCuSe$_2$ (b) under open boundary condition.

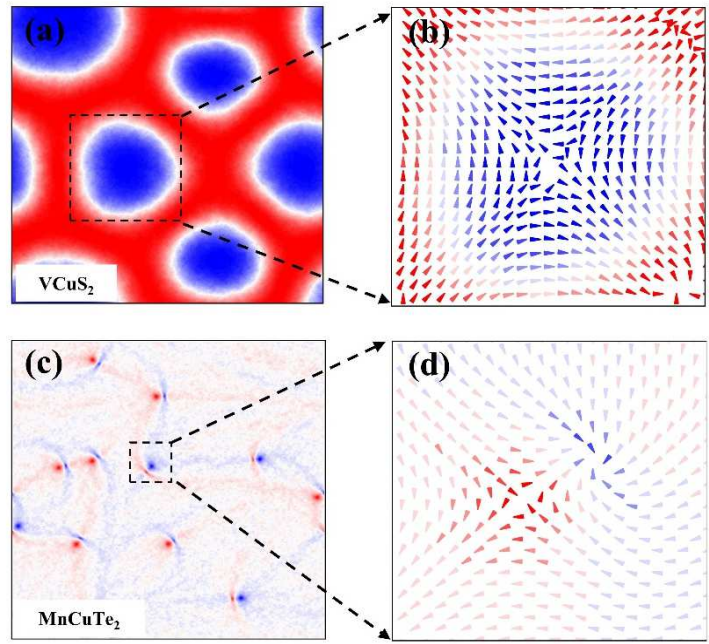

Supplementary Fig. 7. The spin configuration of 100 nm square and zoom of isolated magnetic profile of VCuS$_2$ (a), MnCuTe$_2$ (b) including the dipolar interaction.

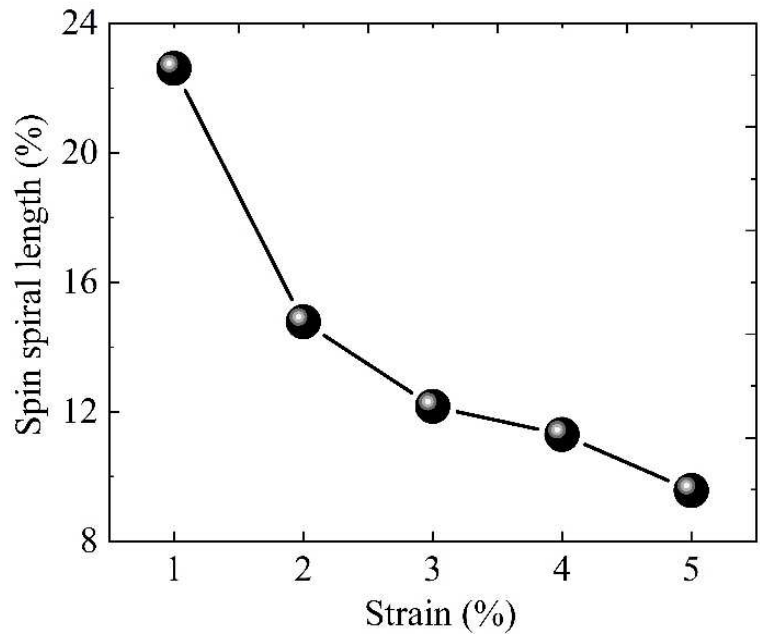

Supplementary Fig. 8. Spin spiral length under increased tensile strain for VCuSe$_2$.

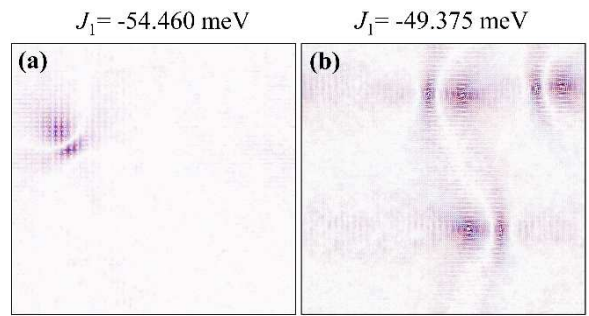

Supplementary Fig. 9. The spin configurations of isolated bimeron (a) and many bimerons (b) in $CoCuS_2$.

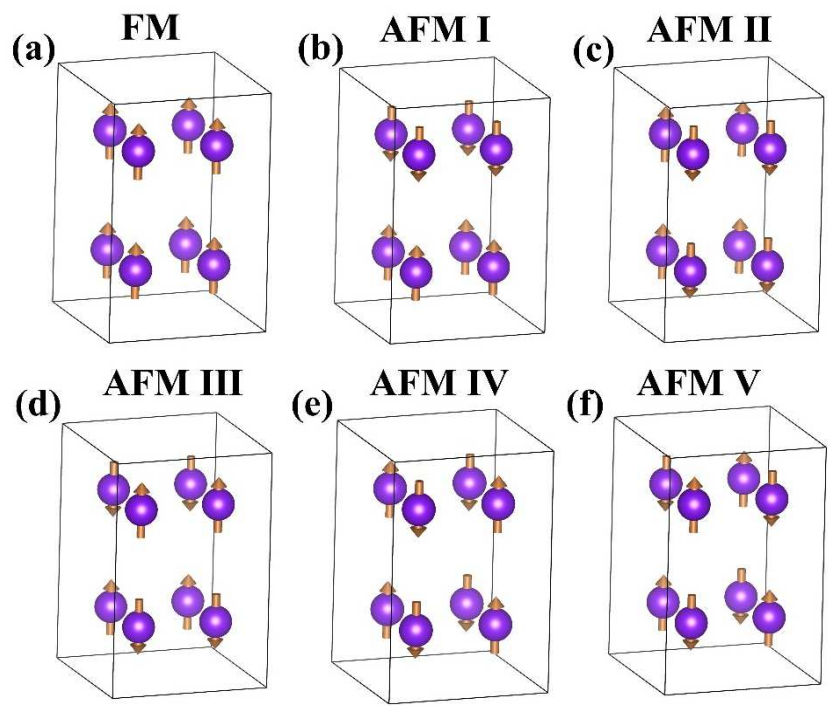

Supplementary Fig. 10. Crystal structure of FeCuTe$_2$ and the six considered magnetic configurations: (a) FM, (b) AFM I, (c) AFM II, (d) AFM III, (e) AFM IV, and (f) AFM V.

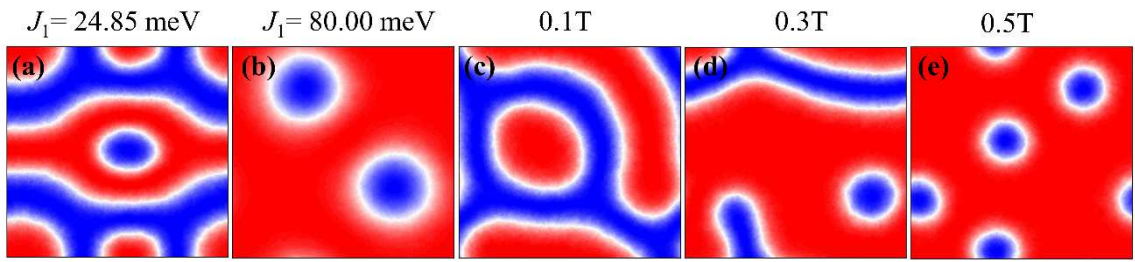

Supplementary Fig. 11. Spin configurations for monolayer VCuS$_2$. Including (a) $J_1$ = 24.85 meV, (b) $J_1$ = 80.00 meV, (c) 0.1 T, (d) 0.3 T and (e) 0.5 T.